\documentclass{aastex62}

\usepackage{graphicx}
\usepackage{bm}
\usepackage{amsmath}
\usepackage{amssymb}
\usepackage{hyperref}

\received{ }
\revised{ }
\accepted{ }

\submitjournal{ApJ}

\shorttitle{Adsorption of Organic Molecules on Onion-Like-Carbons}
\shortauthors{Qi et al.}

\begin{document}

\title{Adsorption of Organic Molecules on Onion-Like-Carbons: Insights on the Formation of Interstellar Hydrocarbons}

\correspondingauthor{Zhao Wang}
\email{zw@gxu.edu.cn}

\author{Haonan Qi}
\affil{Guangxi Key Laboratory for Relativistic Astrophysics, Department of Physics, Guangxi University, Nanning 530004, P. R. China.}

\author{Sylvain Picaud}
\affil{Institut UTINAM, CNRS UMR 6213, Universit\'{e} Bourgogne Franche-Comt\'{e}, 25030 Besan\c{c}on, France.}
	
\author{Michel Devel}
\affil{FEMTO-ST institute, CNRS, ENSMM, 15B avenue des Montboucons, 25030 Besan\c{c}on, France.}
	
\author{Enwei Liang}
\affil{Guangxi Key Laboratory for Relativistic Astrophysics, Department of Physics, Guangxi University, Nanning 530004, P. R. China.}

\author{Zhao Wang}
\affil{Guangxi Key Laboratory for Relativistic Astrophysics, Department of Physics, Guangxi University, Nanning 530004, P. R. China.}

\begin{abstract}
Using atomistic simulations, we characterize the adsorption process of organic molecules on carbon nanoparticles, both of which have been reported to be abundant in the interstellar medium (ISM). It is found that the aromatic organics are adsorbed more readily than the aliphatic ones. This selectivity would favor the formation of polycyclic aromatic hydrocarbons (PAHs) or fullerene-like structures in the ISM due to structural similarity. It is also observed in our simulations that the molecules form a monolayer over the nanoparticle surface before stacking up in aggregates. This suggests a possible layer-by-layer formation process of onion-like nanostructures in the ISM. These findings reveal the possible role of carbon nanoparticles as selective catalysts that could provide reaction substrates for the formation of interstellar PAHs, high-fullerenes and soots from gas-phase molecules.
\end{abstract}


\section{Introduction} \label{sec:intro}

Recent observations \textit{via} the Infrared Space Observatory (ISO) and ground-based radiotelescopes have pointed to the abundance of diverse carbonaceous solids in the interstellar medium (ISM) \citep{Ehrenfreund2000,Pirali2007}. Different forms of cosmic carbon materials have been reported, including fullerenes, diamond, graphite, soot, and so on \citep{Kroto1992,Henning1998,Cami2010}. These carbon solids are likely to play key roles in the physical and chemical evolution of the ISM \citep{Garcia-Hernandez2010,Chen2017,Krasnokutski2017}. Besides the carbon solids, organic molecules have also been proven to be abundant in the ISM, as evidenced by recent space and ground observations \citep{Muller2001,Kaiser2002,Herbst2009,Belloche2013,Belloche2014,Etim2016,Ceccarelli2017,Weaver2017} as well as by laboratory investigations on meteoritic samples \citep{Henning1998,Geppert2006,Rotelli2016}. The aggregation of small organic molecules is hypothesized to provide a source for the formation of complex organic molecules including amino acids in the primitive Earth atmosphere \citep{Miller1959}, while the processes included in the evolution of the ISM \citep{Jones2011,Chiar2013,Scoville2017} that led to the origin of life are still in a puzzle \citep{Ehrenfreund2006,Piani2017,Puzzarini2017}.

Gas-phase molecules are suspected to be elemental building blocks of complex interstellar organic molecules, many of which are polycyclic aromatic hydrocarbons (PAHs) \citep{Tielens2008}. Interstellar PAHs carry a number of broad infrared emission features, e.g. the $2175\;\mathrm{\AA}$ UV extinction \citep{Blasberger2017}. Fullerenes of various sizes are diffuse interstellar bands (DIBs) carriers \citep{Iglesias-Groth2007,Salama2011,Campbell2015,Omont2016}. The presence of PAHs and carbon nanoparticles has been invoked to explain the atmospheric composition of exoplanets such as the warm Jupiter HD 189733b \citep{Mousis2011}. Moreover, analysis on Allende meteorites by laser desorption mass spectrometry (LDMS) and microscopy experiments has shown onion-like structures of higher fullerenes \citep{Rotundi1998,Becker1999,Jager2006,Krasnokutski2017}. The presence of grain layered carbon nanostructures in the ISM has therefore been unambiguously proposed in many studies \citep{Chiar2013,Shi2015,Anderson2017}. Meanwhile, very little is known about the interaction between these carbon nanostructures and surrounding molecules \citep{Demarais2012}, and about their coupled roles in the evolution of the ISM.

Following the development of microscopic sensing devices, the adsorption of small molecules on carbon nanotubes and graphene has been intensively studied \citep{Liu2017}. An interesting feature is that usually the adsorption enthalpy increases and then decreases with increasing molecule coverage \citep{Lazar2013}. In our previous works, the interaction between water molecules and carbon soots was theoretically investigated on grounds of astronomical and environmental interests \citep{Moulin2007,Hantal2010,Mousis2011}, and molecule aggregation was found to depend on various parameters of the adsorption environment. Molecular adsorption has also been shown to modify the optical properties of soot primary nanoparticles \citep{Xue2009,Fan2016}. However, these results are revealing, the interaction between organic molecules and layered carbon solids remains largely unexplored. Thus, in the present work we characterize the adsorption of organic molecules on layered carbon nanoparticles using atomistic simulations. We focus on the effect of molecule type and density on the aggregation process around the nanoparticle and determine the corresponding adsorption energy.

\section{Methods} \label{sec:method}

\begin{figure}[htp]
\centerline{\includegraphics[width=9cm]{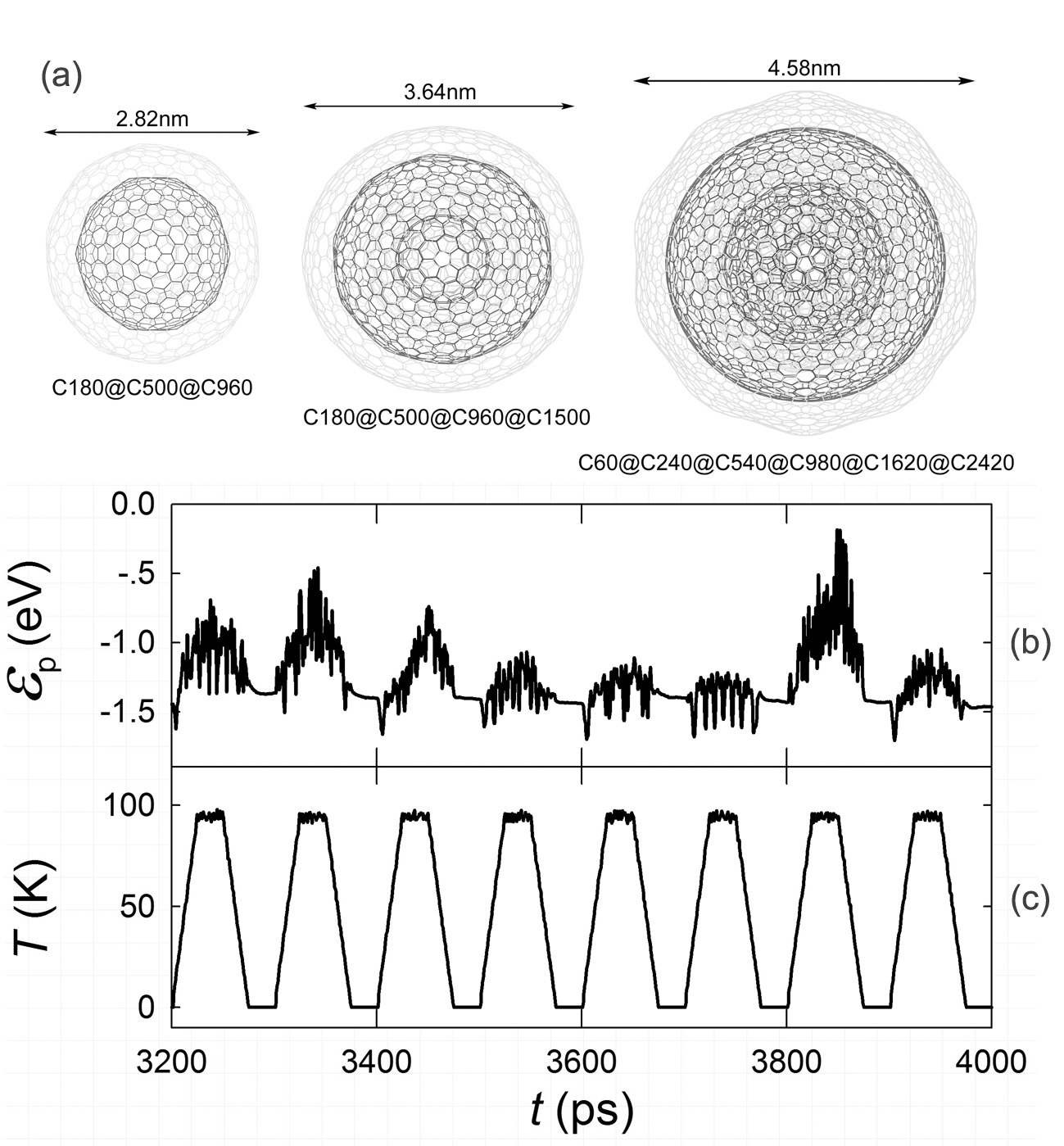}}
\caption{\label{F1}
(a) Atomistic configurations of three ``Russian doll'' carbon nanoparticles studied in this work. (b-c) Variation of the potential energy $\varepsilon_{p}$ and temperature $T$ of a sample during a simulation for instance. The zero-potential value of $\varepsilon_{p}$ is set to be -21767.0 eV.
}
\end{figure}

More than 200 different molecules have been detected in the ISM. Although many oxygen-containing organics have been detected in ISM \citep{Weaver2017}, the present work focuses on carbohydrates for simplicity. Here, two aromatic (Benzene C$_{6}$H$_{6}$ and Toluene C$_{7}$H$_{8}$), two aliphatic linear (Methylpropane C$_{3}$H$_{8}$ and Hexane C$_{6}$H$_{14}$) and one aliphatic cyclic (Cyclopentane C$_{5}$H$_{10}$) organic molecules are chosen as the adsorbates in this work. The adsorption of these molecules on the carbon nanoparticles is simulated by using the parallel computing package LAMMPS \citep{Plimpton1995}. The carbon nanoparticles are modeled by carbon bucky-onions containing three, four or six carbon layers (i.e., fullerenes of increasing size) arranged in a concentric way, as shown in Figure \ref{F1}(a). The spherical fullerene are constructed by introducing Stone-Wales defects on the edges or medians of each triangle containing a pentagon at its vertices of the icosahedral or triacontahedral fullerenes using the method introduced in Ref. \citep{Langlet2007}. These generated structures of fullerenes are optimized before interacting with organics. These $sp^{2}$-hybridized carbon structures are consistent with the observations of onion-like carbons by LDMS \citep{Becker1999} and laboratory experiments \citep{Rotundi1998,Jager2006}. Note that the size of nanoparticle is smaller in the simulations due to limitation of the computational resource. The atomistic interactions between the carbon and hydrogen atoms are described in the framework of the adaptive interatomic reactive empirical bond order (AIREBO) potential \citep{Stuart2000}, in which the total interatomic potential $\varepsilon_{p}$ involves many-body terms as a collection of those of individual bonds,

\begin{equation}
\label{eq1}
\varepsilon_{p}=\frac{1}{2}\sum\limits_{i=1}^N{\sum\limits_{\substack{j=1 \\ j\ne i} }^N{\left[
\begin{array}{l}
\varphi^R\left(r_{i,j}\right)
+b_{i,j}\varphi^A\left(r_{i,j}\right)
+\varphi^{LJ}\left(r_{i,j}\right)
+\sum\limits_{\substack{k=1 \\ k\ne i,j} }^N
{\sum\limits_{\substack{\ell=1 \\ \ell\ne i,j,k}}^N
{\varphi_{kij\ell}^{tor}}}
\end {array}
\right]} }
\end{equation}

\noindent where $\varphi^R$ and $\varphi^A$ model the interatomic repulsion and attraction terms between the valence electrons, respectively. $\varphi^{tor}$ represents the effect of single-bond torsion. The many-body effect is included in the bond order function,

\begin{equation}
\label{eq3}
b_{ij} = \frac{1}{2} \left( b_{ij}^{\sigma-\pi} + b_{ji}^{\sigma-\pi} + b_{ji}^{RC}+ b_{ji}^{DH} \right)
\end{equation}

\noindent where $b_{ij}^{\sigma-\pi}$ depends on the atomic distance and bond angle, $b_{ji}^{RC}$ represents the bond conjugation effect, and $b_{ji}^{DH}$ is a dihedral-angle term for double bonds. The parameterization of this potential is provided elsewhere \citep{Stuart2000}. The long-range interactions are included by adding $\varphi^{LJ}$, a parametrized \textit{Lennard-Jones} (LJ) potential with a cutoff radius of $1.0\;\mathrm{nm}$. Note that $\varphi^{LJ}$ applies to both the inter- and intra-molecule atomistic pairwise interactions in order to enable a smooth transition between the long-range and covalent interactions. Comparing to other force fields for hydrocarbon systems, the modeling of the bond rotation and torsion of the AIREBO potential with respect to the bond order are particularly important for simulating the adsorption process, taking into account possible deformation of the substrate induced by the adsorbate and \textit{vice versa}. The AIREBO potential has therefore recently shown good accuracy in describing adsorption behaviors of systems containing $sp^{2}$- and $sp^{3}$-hybridized carbons \citep{Petucci2013,Sun2017}.

The molecules could be adsorbed at random sites on the nanoparticle surface in a real molecular cloud. In such a process, low-energy states would finally dominate the statistics. It is thus crucial to find the possible low-energy ground states of the adsorption system. To this end, we raise artificially the temperature to transit to other metastable states. The organic molecules are initially placed at random sites on the surface of the carbon nanoparticles. The equilibrium configuration of adsorption is computed by minimizing the total potential energy $\varepsilon_{p}$ of the system \citep{Wang2009a,Wang2009b,Wang2007a,Wang2007b}. A set of molecular dynamics (MD) \citep{Guo2015,Lin2014,Wang2011} are then performed in order to let molecules move randomly on the surface by including a repeated annealing process, during which the temperature of the system $T$ is controlled to switch between $\sim 0$ and $100\;\mathrm{K}$, and the potential energy $\varepsilon_{p}$ therefore circles with the variation in $T$, as illustrated in Figure \ref{F1}(b-c). Note that the temperature fluctuation here does not have realistic astrophysical sense, since the time scale in the MD simulations is not comparable to that involved in the ISM evolution. Note, that there is still a probability of molecule redistribution enabled by  local temperature fluctuation, since the lifetime of interstellar molecular cloud is at the order of million years \citep{Elmegreen2007}.

The equations of motion are integrated by the Verlet algorithm with a time step of $0.5\;\mathrm{fs}$. Despite below $100\;\mathrm{K}$ the dynamics could be possibly dominated by ground-state motion that cannot be accurately reproduced with classical simulations, MD are used to perturb the system by adding thermal fluctuation before new ground-state configurations of adsorption can be generated by subsequent structural optimization. Given that the temperature of the ISM could be as low as a few Kelvins \citep{Lippok2016}, the adsorption energy $\varepsilon_{mol}$ corresponding to a ground-state ($\sim 0\;\mathrm{K}$) configuration is defined as

\begin{equation}
\label{eq4}
\varepsilon_{mol} = \frac{\varepsilon_{p} - \varepsilon_{np} - \varepsilon_{ads}}{N},
\end{equation}

\noindent where $\varepsilon_{np}$ stands for the potential of the nanoparticle when the small molecules are instantly removed from the surface, $\varepsilon_{ads}$ is that of all organic molecules together without the nanoparticle, and $N$ is the total number of the molecules. By this definition, $\varepsilon_{mol}$ stands for the energy of the long-range (van der Waals) interaction between the nanoparticle and molecules, since the chemisorption is not considered in the present work due to that only ground-states are taking into account with no initially non-saturated bond. For statistical relevance, this energy is averaged over $20$ different ground-state configurations for each simulation. The number $20$ has been chosen according to the results of a numerical test.

\section{Results and discussions} \label{sec:RandD}

\begin{figure}[htp]
\centerline{\includegraphics[width=9cm]{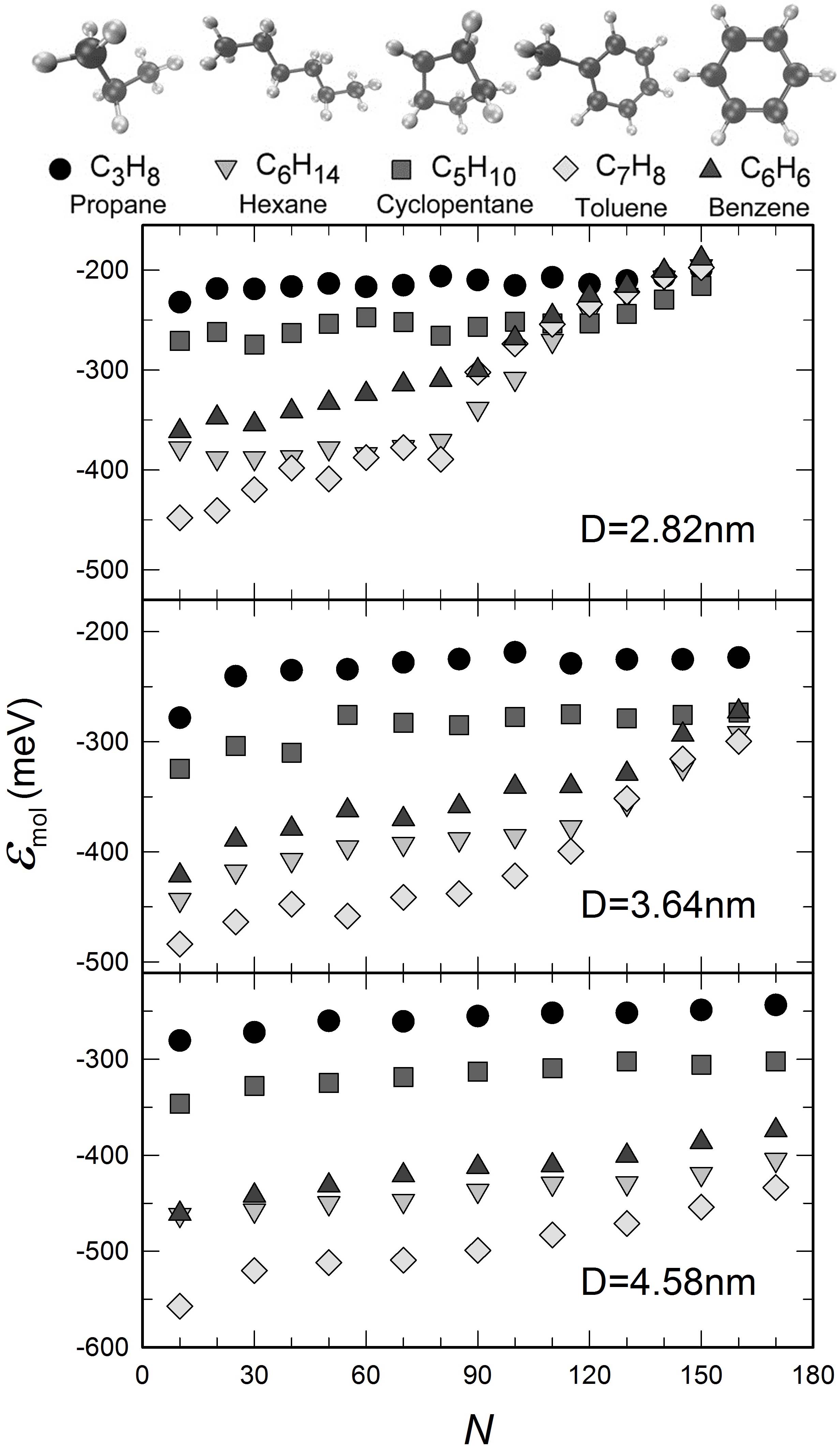}}
\caption{\label{F2}
Adsorption energy per molecule versus the number N of five different organic molecules adsorbed on three nanoparticles of different sizes.
}
\end{figure}

Figure \ref{F2} shows the adsorption energy per molecule $\varepsilon_{mol}$ as a function of the number of adsorbed molecules $N$ for the five types of organic molecules on three carbon nanoparticle of different sizes considered here. It is shown that $\varepsilon_{mol}$ first increases with increasing $N$ before saturating at a certain value, and then increases again with $N$. This behavior is more significant for the relatively large molecules C$_{6}$H$_{14}$ and C$_{7}$H$_{8}$ than for the smaller ones, and is due to a surface-coverage effect, to be discussed below. The comparison between $\varepsilon_{mol}$ for different adsorbed molecules shows that the absolute values of $\varepsilon_{mol}$ is larger for the molecules with more carbon atoms, while the effect of hydrogen seems to be less significant. For instance, the toluene (C$_{7}$H$_{8}$), which has the largest number of carbon atoms, is most-strongly adsorbed to the surface among the species investigated here, while $\left| \varepsilon_{mol} \right|$ for the hexanes (C$_{6}$H$_{14}$) are found to the slightly higher than that for the benzenes (C$_{6}$H$_{6}$). This is because the interactions is dominated by van der Waals interactions, while the L-J potential well for C-H is only 0.5523 of that for C-C interaction \citep{Stuart2000}. Note results of an alternative definition of the adsorption energy is provided in APPENDIX \ref{appendix1}

$\varepsilon_{mol}$  includes the effects of both the atom number and molecule geometry. To discriminate more clearly between these two effects,  we define an adsorption energy per atom as

\begin{equation}
\label{eq4b}
\varepsilon_{atom} = \frac{\varepsilon_{mol}}{n_{C}+\xi n_{H}}
\end{equation}

\noindent where $n_{C}$ and $n_{H}$ are the numbers of carbon and hydrogen atoms in the molecule, respectively. The parameter $\xi$ is taken to be $0.5523$, the ratio between the C-H and C-C equilibrium interaction potentials given in the parameters for the LJ function \citep{Stuart2000}.

\begin{figure}[htp]
\centerline{\includegraphics[width=9cm]{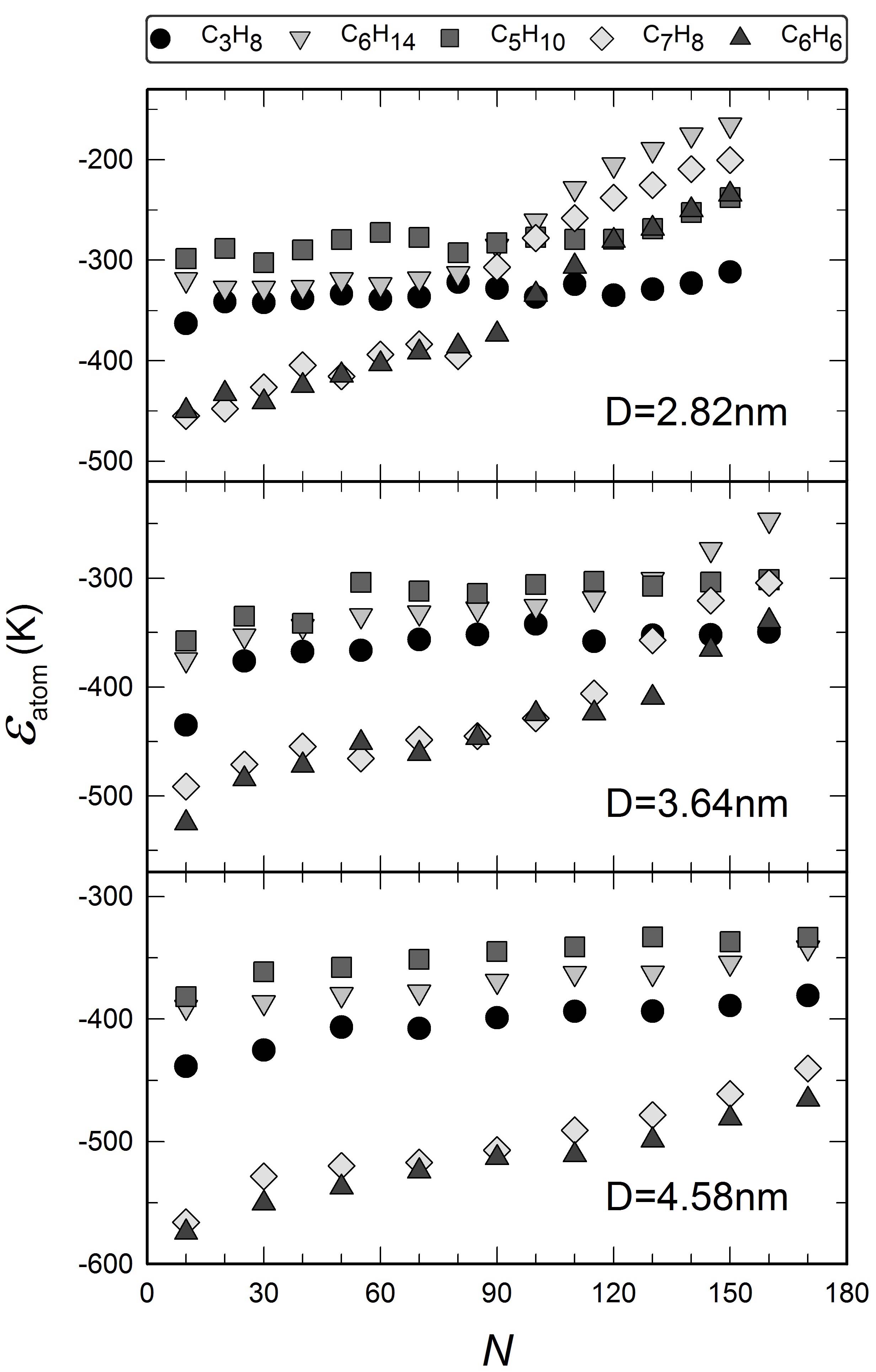}}
\caption{\label{F3}
Adsorption energy per atom (in Kelvin) \textit{versus} the number $N$ of organic molecules for the five molecule types on three nanoparticles of different sizes.
}
\end{figure}

We plot in Figure 3 $\varepsilon_{atom}$ as a function of N to highlight the effect of molecule shape on the adsorption energy. It can be seen that the aromatic molecules C$_{6}$H$_{6}$ and C$_{7}$H$_{8}$ exhibit high absolute values of the adsorption energy, while that of the linear molecules C$_{3}$H$_{8}$ and C$_{6}$H$_{14}$ are characterized to be lower. The cyclopentane (C$_{5}$H$_{10}$) is found to have the lowest absolute values of adsorption energy. As the number of the molecule increases beyond a certain threshold, the slope of energy curve changes due to aggregation of the molecules as we are going to discuss below. Since the energy is calculated for physisorption in this work, the energy magnitudes are generally smaller than those for chemical adsorption \citep{Das2018, Hama2013, Wakelam2017}.

\begin{figure}[htp]
\centerline{\includegraphics[width=9cm]{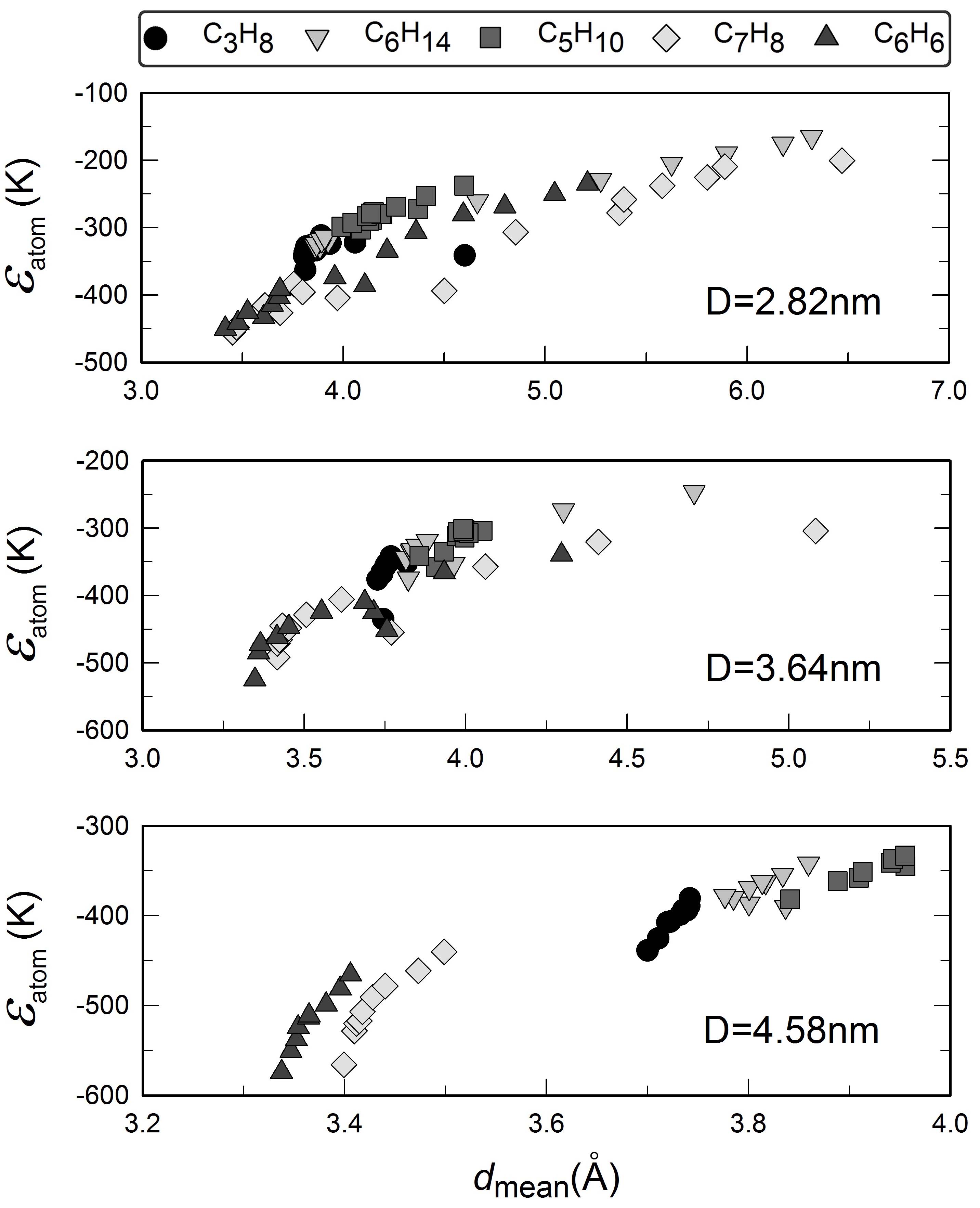}}
\caption{\label{F4}
 Adsorption energy per atom \textit{versus} mean atomistic adsorbate-substrate spacing for various molecule numbers on three nanoparticles of different sizes.
}
\end{figure}

The values of the adsorption energy are found to be strongly correlated to the average interatomic spacing $d_{mean}$ between the molecules and nanoparticle surface, as shown in Figure \ref{F4}. We can see that the per-atom adsorption energy increases with increasing mean atomistic adsorbate-substrate spacing in a power-law manner, as determined by the $LJ$ interaction potential function \citep{Jones1924}. The values of the atomistic adsorbate-substrate spacing is also plotted in APPENDIX \ref{appendix2} as functions of the number $N$. The differences in the values of $d_{mean}$ could be related to the contact commensurability between the lattices of adsorbate and substrate \citep{Verhoeven2004}, which seems to favor the adsorption of aromatic (hexagonal) molecules by a preferred way of $\pi-\pi$ stacking. It is also marked that the values of $d_{mean}$ calculated for the hexane (C$_{6}$H$_{14}$) first decreases when $N$ starts to increase. This trend is due to the bending of the carbon chain, as shown by the inset of Figure \ref{F9}, which thus appears to be as a particular adsorption feature of the linear molecules. This feature highlights the influence of the structural flexibility on the adsorption of chain-shaped molecules is highlighted.

\begin{figure}[htp]
\centerline{\includegraphics[width=9cm]{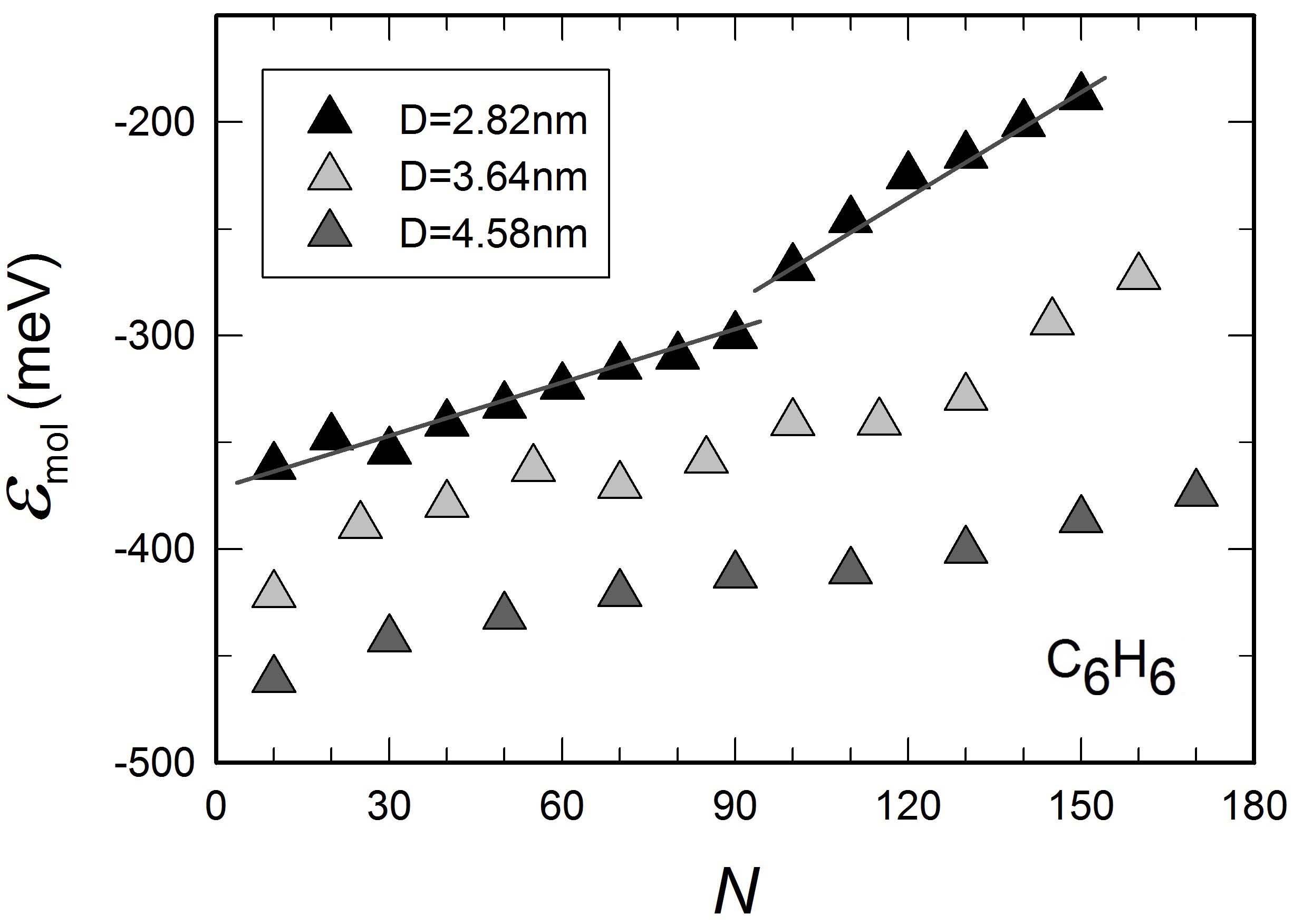}}
\caption{\label{F5}
Adsorption energy per molecule as a function of the number $N$ of C$_{6}$H$_{6}$ adsorbed on three nanoparticles of different sizes.
}
\end{figure}

 The reaction process for the formation of the PAHs from organic molecules is an open question. Benzene molecules are suspected to be the building blocks of PAHs in the ISM \citep{Joblin2018}. To reveal the role of the carbon solid in such a transformation, we plot the adsorption energy of benzene molecules on three nanoparticles of different sizes in Figure \ref{F5}. It can be seen that the adsorption energy per molecule $\varepsilon_{mol}$ decreases as the nanoparticle size increases. This is expected since a larger nanoparticle will have a flatter surface which is supposed to favor the adsorption. More importantly, it is shown by the different slopes of the eye-guide lines in Figure \ref{F5} that the molecules interact stronger with the nanoparticle than with other molecules. This implies that the energy-optimized stacking manner of the molecules exhibits a 2D manner, i.e., on the nanoparticle surface, it is energetically favorable that molecules form a thin layers instead to stack up in three-dimensional aggregate. This is double-confirmed by computing the potential energy of molecule-molecule interactions as shown by the table in APPENDIX \ref{appendix3}. The energy magnitude of the molecule-molecule interaction is smaller than that of the molecules-nanoparticle interaction shown in Figure \ref{F2}. Moreover, it can be seen that the behavior of the adsorption energy is strongly correlated with the surface coverage, in particular, for the smallest nanoparticle with $D=2.82\;\mathrm{nm}$.

\begin{figure}[htp]
\centerline{\includegraphics[width=13cm]{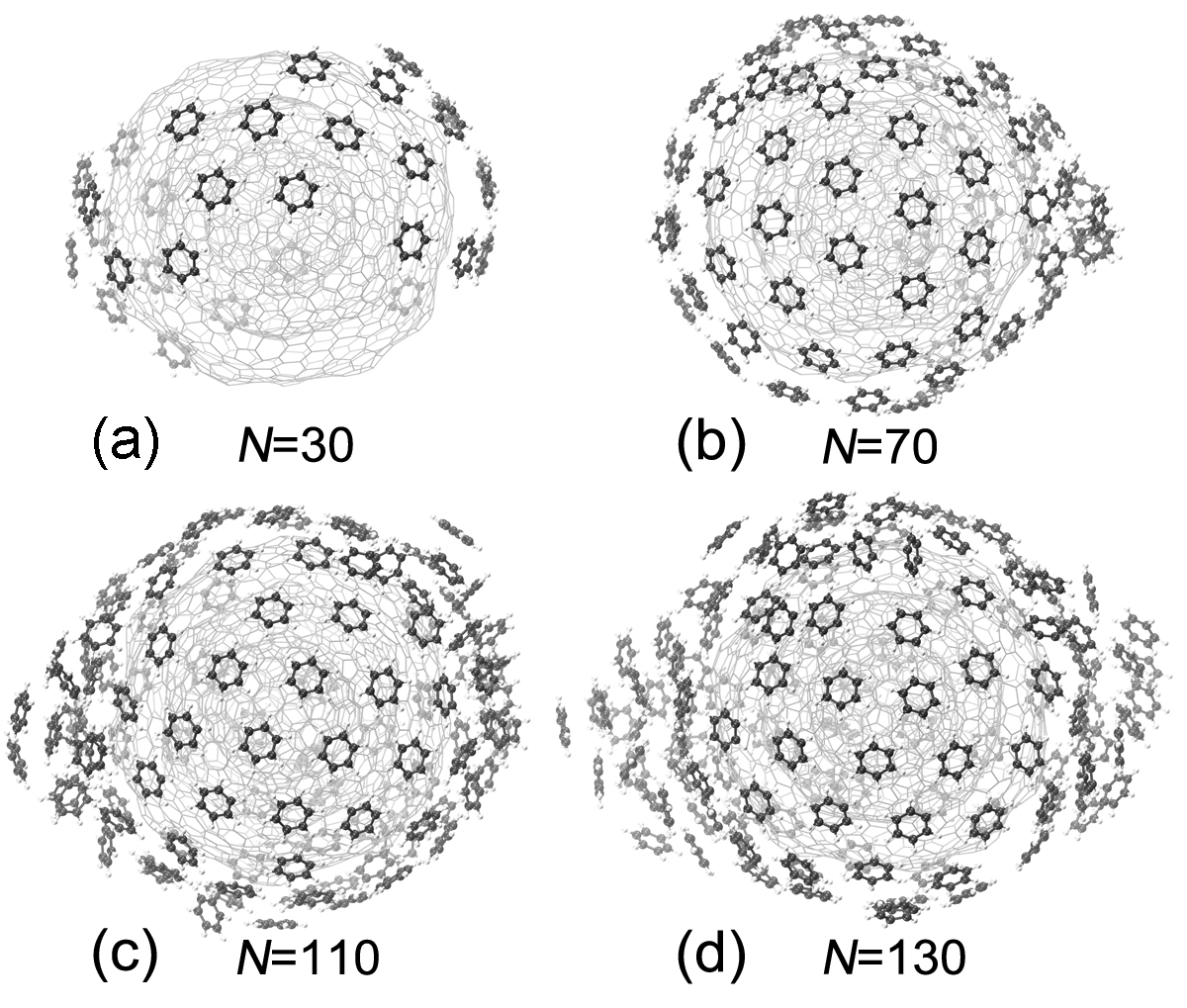}}
\caption{\label{F6}
Atomistic configurations of the adsorbed benzene molecules on a carbon nanoparticle $D=2.82\;\mathrm{nm}$ with different surface coverages with (a) $N=30$,(b) $N=70$,(c) $N=110$, (d) $N=130$. The carbon and hydrogen atoms in benzenes are represented by black and white spheres, respectively.
}
\end{figure}

Indeed, when a small number of molecules are adsorbed, they are observed to distribute homogeneously on the nanoparticle surface forming a thin monolayer, as shown in Figure \ref{F6} (a) and (b). The absolute value of the adsorption energy is roughly a linear function of the number of adsorbed molecules, before the surface is saturated (Figure \ref{F5}). After the nanoparticle surface is about to be fully covered with increasing adsorbed molecules, added molecules start to stack up to form three-dimensional aggregates, as shown in Figure \ref{F6} (c) and (d). The progressive stacking of molecules leads to another linear increase of the adsorption energy, with however lower energy magnitude. By considering that a benzene molecule has roughly a radius of $0.263\;\mathrm{nm}$, and by taking into account the LJ parameter of the benzene-benzene interaction model, we shall estimate that each benzene molecule roughly covers a surface of $0.35\;\mathrm{nm}^{2}$. Given that the calculations show that benzene molecules are adsorbed at a distance of about $0.339\;\mathrm{nm}$ to the particle surface, the surface area of the benzene layer directly in contact with the nanoparticle can be estimated to be equal to about $38$, $58$ or $87\;\mathrm{nm}^{2}$ for the particle of a diameter of $2.82$, $3.64$ or $4.58\;\mathrm{nm}$, respectively. This leads to a rough estimation of the maximum number of the benzene molecules that can be packed in a homogeneously saturated ad-layer that is in direct contact with the nanoparticle, equals to about $108$, $165$ or $248$ molecules for the nanoparticle of a diameter of $2.82$, $3.64$ and $4.58\;\mathrm{nm}$, respectively.

Thus, the maximum number of molecules considered in the present simulations ($180$ benzene molecules) is \textit{a priori} not sufficiently large to fully saturate the first layer of the largest nanoparticle simulated here. This could explain why $\varepsilon_{mol}$ increases smoothly with $N$ for the largest nanoparticle, without experiencing the slope break that is observed for the smallest nanoparticle. This energetic difference highlights the difference between the molecule-nanoparticle and molecule-molecule adsorptions, and leads to a hypothesis on the PAH processing in interstellar molecular clouds as suggested below. Moreover, we provide a number of atomistic configurations of the adsorbed molecules on different nanoparticles (see Appendix \ref{appendix4}), which will be useful for investigations on optical properties of carbonaceous structures in the ISM.

\section{Conclusion} \label{sec:concl}

\begin{figure}[htp]
\centerline{\includegraphics[width=16cm]{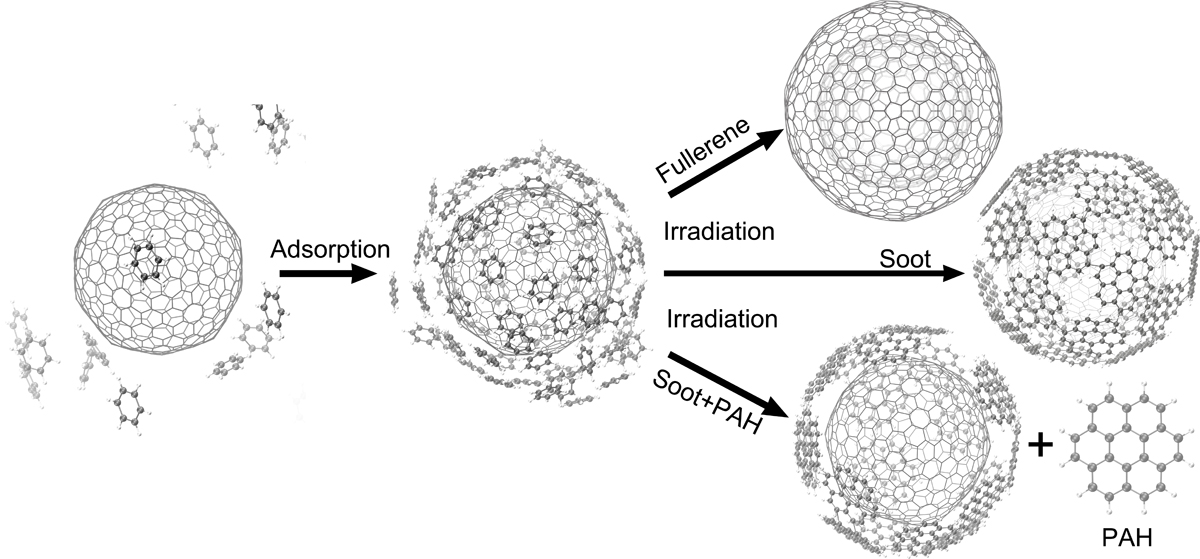}}
\caption{\label{F7}
Schematics of a hypothesis on the formation processes of the high-fullerene, soot and PAHs from organic molecules adsorbed on a nanoparticle in the ISM.
}
\end{figure}

In conclusions, the adsorption of various organics on carbon nanoparticles of different sizes is investigated by atomistic simulations. Our results show that the aromatic molecules exhibit stronger interactions with the nanoparticles than the aliphatic ones. Thus, in presence of possible external stimulation and consequent thermal fluctuation in the ISM environment, this selectivity may favor the formation of $sp^{2}$-hybridized carbon nanostructures such as PAHs or fullerene-like layers due to similarity in their atomic structure and bond conjugation. Note that the temperature variation technique employed in the simulations is unrelated to the aforementioned thermal fluctuation in the ISM. It is also observed in our simulations that the molecules form spontaneously a homogenous layer over the nanoparticle surface before stacking up in 3D aggregations. This indicates a possible layer-by-layer formation process of onion-like nanostructures in the ISM. This is also based on the fact that the interaction energy is lower in absolute value between molecules in aggregations than between molecules and nanoparticles. Since the time scale for in the ISM evolution is so long \citep{Elmegreen2007} that redistribution of molecules over the nanoparticle surface could be enabled by external stimulation or local temperature fluctuation, low-energy states would finally dominate the statistics. Hence, with the energy input from the expected stimulation processes into the nanoparticle or molecules, the energy difference between molecule-molecule and molecule-nanoparticle interactions can lead to selectively disperse aggregations, while keeping a homogenous layer over the nanoparticle surface.

The above results demonstrated a physisorption process in the ISM as the first step of chemisorption, and lead to hypotheses illustrated in Figure \ref{F7} on the formation of layered carbon nanoparticles and PAHs from elementary molecules in the ISM. In such a scheme, gas-phase molecules, preferably aromatic, are adsorbed on the surface of small nanoparticles, as evidenced in the present theoretical study. Irradiation [by cosmic-ray particles, UV photons, etc. \citep{Chiar2013, Zheng2008}] would lead to the formation of various possible nanostructures depending on the magnitude of dehydrogenation: e.g., onion-like fullerenes, PAHs or soots (\textit{a priori} PAH clusters). We should note that the hypotheses illustrated in Figure 7 are idealized. In real ISM environment, the accretion of other abundant gaseous species including aliphatic molecules would inhibit the successive assembly of aromatic molecules, despite aromatic molecules are adsorbed stronger than other ones. Thus, the surface density of aromatic and aliphatic molecules should depend on their gas-phase abundances. These findings have implications on the formation of layered carbon nanoparticles and PAHs in the ISM (mostly in the environment of diffuse clouds) from elementary gas-phase molecules. In this scenario, possible catalyst role of the nanoparticles is hypothesized for the molecular complexity in the ISM, as an initial aspect in the nucleation and growth of interstellar dust and in star and planet formation.

\acknowledgments
Jes\'{u}s Carrete is acknowledged for proof-reading of this manuscript. This work has been supported by the Guangxi Science Foundation (2018GXNSFAA138179, 2013GXNSFFA019001), Scientific Research Foundation of Guangxi University (XTZ160532), Guangxi Key Laboratory Foundation (15-140-54), National Natural Science Foundation of China (11533003 and U1731239), EIPHI Graduate School (ANR-17-EURE-0002) and Region Bourgogne Franche-Comt\'{e}.

\appendix

\section{An alternative adsorption energy} \label{appendix1}
Alternatively, the adsorption energy can be defined as the amount of energy needed to bring together the nanoparticle and molecules from infinite-far.

\begin{equation}
\label{eq5}
\varepsilon^{*}= \varepsilon_{total} - \varepsilon^{*}_{np} - N \varepsilon^{*}_{ads}
\end{equation}

\noindent where $\varepsilon^{*}_{np}$ is the minimized potential energy of the nanoparticle, and $\varepsilon^{*}_{ads}$ is that of a single adsorbate.

\begin{figure}[htp]
\centerline{\includegraphics[width=8cm]{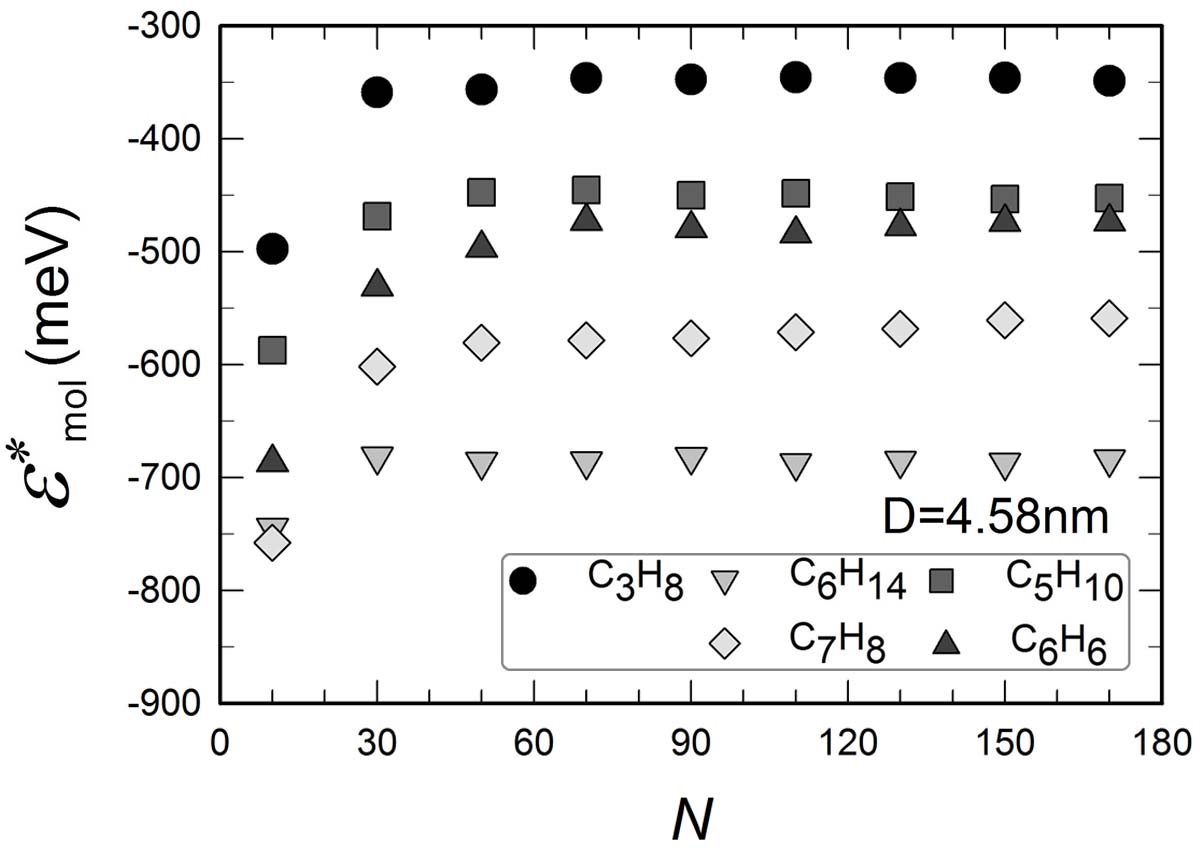}}
\caption{\label{F8}
Adsorption energy per molecule \textit{versus} the number $N$ of organic molecules adsorbed on a  nanoparticle of $4.58$ nm in diameter.
}
\end{figure}

\section{Atomistic adsorbate-substrate spacing} \label{appendix2}
Mean atomistic adsorbate-substrate spacing as functions of the number $N$ of organic molecules.

\begin{figure}[htp]
\centerline{\includegraphics[width=8cm]{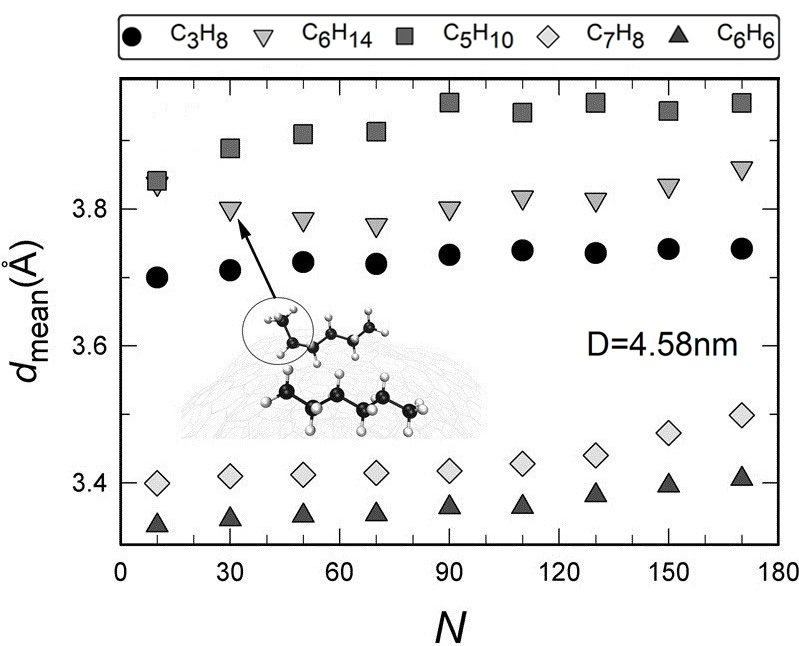}}
\caption{\label{F9}
Mean atomistic adsorbate-substrate spacing as functions of the number $N$ of organic molecules for the nanoparticle $D=4.58\,$nm.
}
\end{figure}

\section{Molecule-molecule interaction energy} \label{appendix3}
Table \ref{tabapp} gives the molecule-molecule interaction energy that has been calculated by using 12 different equilibrium configurations of two molecules together. $\varepsilon_{min}$ and $\varepsilon_{max}$ give the minimum and maximum values in the $20$ samples with various initial configurations, respectively. $\varepsilon_{ave}$ gives the mean.

\begin{deluxetable}{cccc cccc}
\tablecaption{Molecule-molecule interaction energy values \label{tabapp}. }
\setlength{\tabcolsep}{6mm}{
\tablehead{\colhead{$\varepsilon_{mol}$ (meV)} & \colhead{C$_{6}$H$_{6}$} & \colhead{C$_{5}$H$_{10}$} & \colhead{C$_{6}$H$_{14}$} & \colhead{C$_{3}$H$_{8}$} & \colhead{C$_{7}$H$_{8}$}}
\startdata
$\varepsilon_{min}$ & -17.02 & -63.91 & -33.34 & -36.89 & -27.06 \\
$\varepsilon_{max}$ & -154.06 & -12.171 & -166.58 & -80.17 & -212.21 \\
$\varepsilon_{ave}$ & -81.71& -90.22 & -116.46	& -61.48 & -158.67
\enddata}
\end{deluxetable}

\section{Atomistic coordinate data files} \label{appendix4}

Data files that contain optimized atomistic configurations of organic molecules adsorbed on onion-like nanoparticles are provided on the AAS website. These .xyz files contain the atomistic coordinates of the adsorbed organic molecules and nanoparticles by giving the total number of atoms that will be read on the first line, the numbers of molecules, atoms in each molecule, and atoms in the nanoparticle on the second, and the atom type and three atomic Cartesian coordinates in the following lines. Please see the PDF in the zipped Enhanced Manuscript Information file for a complete list and instructions.

%\bibliography{refs}

\end{document}